\documentclass[12pt]{iopart}
\usepackage{graphics}
\usepackage{epsfig}
\usepackage{color}
\usepackage{colordvi}
\newcommand{\prl}{Phys. Rev. Lett. }
\newcommand{\pra}{Phys. Rev. A }
\newcommand{\pre}{Phys. Rev. E }

\begin{document}

\title{Solitons in dipolar Bose-Einstein condensates  with trap and barrier potential}
\author{F. Kh. Abdullaev$^{1,2}$, V.A. Brazhnyi$^3$}
\address{$^1$ Physical - Technical Institute, Uzbek Academy of
Sciences, 2-b, G. Mavlyanov str., 100084, Tashkent, Uzbekistan; \\
$^2$Centro de F\'{\i}sica Te\'orica e Computacional, Faculdade de Ciencias, Universidade de Lisboa, Lisboa 1649-003, Portugal \\
$^3$ Centro de F\'{\i}sica do Porto, Faculdade de Ci\^encias, Universidade do Porto, R. Campo Alegre 687, Porto 4169-007, Portugal}

\date{\today}

\begin{abstract}
The propagation of solitons in dipolar BEC in a trap potential with a barrier potential is investigated.
The regimes of soliton transmission, reflection and splitting  as a function of the ratio between the local and dipolar nonlocal interactions are analyzed analytically and numerically.
{
Coherent splitting and fusion of the soliton by the defect is observed.}
The conditions for  fusion of splitted solitons are found. In addition the delocalization transition governed by the strength of the nonlocal dipolar interaction is presented. Predicted phenomena can be useful for the design of a matter wave splitter and interferometers using matter wave solitons.
\end{abstract}
\maketitle

\section{Introduction}

The Bose-Einstein condensate (BEC) of chromium ($^{52}$Cr), where
long--range dipolar interaction between atoms plays the dominant
role, is a novel kind of nonlinear system becoming available to experiments \cite{griesmaier}.
Properties of dipole-dipole (DD) interactions, namely their long--range character and anisotropy, {allow} dipolar condensates to exhibit many unusual properties not found in BECs with just contact interactions \cite{Lahaye,Baranov}. In particular the existence
of stable isotropic and anisotropic two--dimensional (2D) solitons has been predicted for such cold quantum gases \cite{tmv2008,Eichler}.
Recently the bright solitons in quasi-1D dipolar BEC with competing local and nonlocal interactions  have been studied in works \cite{sinha,Young,cuevas}.

The long-range dipolar interactions become dominant when the local part is detuned to zero by the Feshbach resonance (FR)
techniques, as in the experiment on observation of Anderson localization in {non-interacting cold quantum} gases \cite{Hulet2010}.
In this particular case the pure dipolar bright soliton can be observed.
The propagation of such solitons under joint  action of the trap and a barrier potential, including processes of crossing the barrier by the soliton during its oscillations in the trap are of the greatest interest to investigate.
Such processes have a fundamental importance for the problems of entanglement of quantum solitons in cold dipolar gases.
An interesting limit is the case of strong nonlocality when the dynamics of wavepackets become almost linear.

This paper is devoted to the
investigation of
the properties of cold quantum gases in the presence of long--range dipolar interactions at the mean field level.
First we study oscillations of the dipolar solitons in trap potential using the variational approach (VA) and the scattering theory.
 Second we consider the soliton transmission, reflection and splitting  through a barrier placed in the center of the trap.
Particular attention will be devoted to a process of coalescence of colliding wavepackets at the barrier {which could be} important for the design of beamsplitters and matter wave interferometers using matter wave solitons\cite{Dyke}.
Finally, by means of numerical simulations of the original dynamical equation we verify the
predictions of the VA and analyze the results beyond the analytical predictions.

\section{The model}

We consider the quasi-1D dipolar BEC loaded in a parabolic trap with a barrier at the center of the trap. The governing equation is the 1D Gross-Pitaevskii equation (GPE) with a nonlocal interaction term~\cite{sinha,fattori}:
\begin{eqnarray}
& & i\hbar {\partial \Psi\over{\partial T}} + {\hbar^2\over{2m}} {\partial^{2} \Psi\over{\partial X^{2}}} - V_{tr}(X) \Psi - 2\hbar a_s\omega_{\perp}|\Psi|^{2} \Psi  \nonumber\\  
& &- \frac{2\epsilon d^2}{l_{\perp}^3} \Psi(X,T)\ \int_{-\infty}^{\infty} d\xi R(|X-\xi|)\ |\Psi(\xi,T)|^{2} = 0.
\label{e1}
\end{eqnarray}
Here 
$\omega_{\perp}$ corresponds to the transverse trap frequency, $l_{\perp} = \sqrt{\hbar/m\omega_{\perp}}$, and $d$ is the magnetic dipole moment oriented along the $X$-direction.
{
The parameter $\epsilon$ is connected to the angle $\varphi$ which the dipoles form with the longitudinal axis $X$. If dipoles rotate rapidly in the plane perpendicular to the axis $X$, the parameter $\epsilon$ can vary from 1 for dipoles oriented along  the $X$-axis ($\varphi=0$),  to  $-1/2$ in the case of perpendicular orientation to the $X$-axis ($\varphi=\pi/2$) \cite{sinha,Giovanazzi}.}
The wave function is normalized to the number of atoms comprising the BEC, ${\cal N}\equiv \int_{-\infty}^{\infty} |\Psi(X)|^{2} dX$.
The Hamiltonian for our model  has the following form

\begin{eqnarray}
{ H }= \int_{-\infty}^{\infty}dX\Psi^*(X)\left[ -\frac{\hbar^2}{2m}\frac{\partial^2}{\partial X^2} + V_{tr} (X)+ \hbar a_s\omega_{\perp}|\Psi(X)|^2 \right. \nonumber\\
+\left. \frac{\epsilon d^2}{l_{\perp}^3}\int d\xi \Psi^*(\xi)R(X-\xi)\Psi(\xi)\right]\Psi(X).
\end{eqnarray}

Now we define dimensionless parameters:
\begin{eqnarray*}
 t = T \omega_{\perp},\quad x =  X/l_{\perp},\quad g = \frac{\epsilon a_d}{|a_{s0}|},\quad q = \frac{a_s}{|a_{s0}|},\quad \psi = \sqrt{2|a_{s0}|}\Psi,
\end{eqnarray*}
where $a_d = md^{2}/\hbar^2$ is the characteristic scale of the long-range dipolar interactions, and $a_{s0}$ is the background value of the atomic scattering length.
{
The dimensionless s-wave scattering length $q$ is expressed in units of $a_{s0}$ due to the choice of the normalization of the wave function.}
Eq. (\ref{e1}) now  can be written in the dimensionless form  as follows:
\begin{eqnarray} \label{gpe}
&i\psi_t+\frac{1}{2}\psi_{xx}-q|\psi|^2\psi- V_{tr}(x)\psi&\nonumber\\
&-g(t)\psi(x,t)\int_{-\infty }^{+\infty} R(|x-\xi|)\ |\psi (\xi,t)|^2 \ d\xi  = 0&,
\end{eqnarray}%
where $\psi(x,t)$ is the mean-field wave function of the condensate, $q$ is the local contact interaction term,
$g(t)$ is the nonlinear coefficient responsible for long--range dipolar
interactions, assumed to be time-dependent \cite{Giovanazzi}.
The external trapping potential $V_{tr}(x)$ is
\begin{eqnarray}
&V_{tr}(x) = V_{\omega}(x) + V_{d}(x),& \label{V}\\
&V_{\omega}(x)=\frac{1}{2}\omega^2 x^2, \ V_{d}(x)=V_0 e^{-x^2/(2l^2)}.&
\label{Vtrd}
\end{eqnarray}
The frequency $\omega$ is the longitudinal frequency of the trap, $V_0$ and $l$ are the amplitude and the width of the barrier.
Potential $V_d(x)$ for the case of a broad soliton, $l_s \gg l$, can be approximated by a delta-barrier $V_d =\alpha\delta(x)$ with
$\alpha =\sqrt{2\pi l}V_0$.
 The wave function $\psi(x)$ has  normalization  $\int_{-\infty}^{+\infty} |\psi(x)|^2 dx = N$.

The  following two forms for the kernels in the nonlocality term  are possible
\begin{eqnarray}
R_{1}(x) &=& (1+2x^{2}) \, \exp(x^2) \, \mathrm{erfc}(|x|)-2\pi^{-1/2}|x|, \label{R1} \\
R_{2}(x) &=& x_c^3 (x^2+x_c^2)^{-3/2}. \label{R2}
\end{eqnarray}
The former kernel corresponds to the dipolar BEC in a quasi-1D trap
\cite{sinha}, while the latter one, which contains a cutoff parameter
$x_c$ also proposed for dipolar BEC in \cite{cuevas} {(see also \cite{baizakov09})}
is more convenient for analytical treatment.

{
Using the  matching conditions for $R_i(x)$
\begin{equation}
R_{1}(0)=R_{2}(0), \quad \mbox{and} \quad
\int_{-\infty}^{\infty} R_1(x) dx = \int_{-\infty}^{\infty} R_2(x)
dx,
\end{equation}
it can be found that  $x_c = \pi^{-1/2}$. The comparison of profiles of the
kernels (\ref{R1}) and (\ref{R2}) at this choice of $x_c$ shows
very good fit.}
The cutoff parameter $x_c$ has the meaning of an effective size of the dipole {
and the value is fixed by interpolating the function $R_1(x)$ by $R_2(x)$ }.
Actually, it takes the value of the order of the transverse
confinement length, which makes the model one-dimensional, and is
the unit length in Eq. (\ref{gpe}).  In the limit $x \gg x_c$, where DD interaction effects
dominate over the contact interaction effects, both response functions
behave as $ \sim 1/x^3$. Thus, in the following we will use the kernel
function $R_2(x)$ as it is more simpler for analytical treatments in the description of dipolar effects in
BEC, {using for example VA, where the integrals with $R_2(x)$ can be calculated in explicit form}.

\section{Variational approach}

To describe the soliton propagation we employ the VA
\cite{anderson}.
The Lagrangian for Eq.(\ref{gpe}) has the following form:
\begin{eqnarray}
L=\frac{i}{2}(\psi^*\psi_t -c.c.) &-&\frac{1}{2}|\psi_x|^2 - \frac{q}{2}|\psi|^4 -V_{tr}(x)|\psi|^2 \nonumber\\
&-&\frac{g}{2}|\psi|^2\int_{-\infty}^{\infty} R(|x-\xi|)|\psi(\xi,t)|^2 d\xi .
\end{eqnarray}
To derive equations for soliton parameters we use the Gaussian ansatz
\begin{eqnarray}\label{ansatz}
\psi(x,t) &=& A(t)\exp\left\{-\frac{[x-\zeta(t)]^2}{2a^2(t)} + ib(t)[x-\zeta(t)]^2 \right.\nonumber\\
&+& \left. ik(t)[x-\zeta(t)]+i\phi(t)\right\},
\end{eqnarray}
where $A,a,b,\zeta,k,\phi$ are the soliton amplitude, width, chirp, coordinate of the center of mass, {wave vector} and linear phase, respectively.
Calculating the averaged Lagrangian $\bar{L}=\int L dx$ with this ansatz
we obtain
\begin{eqnarray}\label{avL}
\frac{\bar{L}}{N}&=&-\frac{b_t a^2}{2} + k\zeta_t -\phi_t 
-\frac{1}{4a^2} - b^2 a^2 - \frac{k^2}{2} -\frac {1}{2}\omega^2\left(\frac{a^2}{2}+\zeta^2\right)\nonumber\\
&-& \frac{N}{2\sqrt{2\pi}a}\left[q + g F(a)\right] - V_0 G(a,\zeta,l),
\end{eqnarray}
where $N = \sqrt{\pi}A^2 a =$const and
\begin{eqnarray}
F(a)=  \pi^{-1/2}\int_{0}^{\infty}\frac{e^{-\alpha t}}{\sqrt{t}(t+ 1)^{3/2}}dt,\qquad  \alpha = 1/2\pi a^2, \label{F}\\
G(a,\zeta,l)=\frac{\sqrt{2}e^{-\zeta^2/(l^2(2+a^2/l^2))}}{\sqrt{2+a^2/l^2}}.\label{G}
\end{eqnarray}
In computing the integral the shift $x \to x-\zeta$, $\xi \to \xi-\zeta$ and a transformation to the new variables $z=1/2 (x-\xi)$, $y = 1/2(x+\xi)$ and
$dxd\xi = 2dzdy$ have been used. Considering the Euler-Lagrange equations
$$
\frac{\partial (\bar{L}/N)}{\partial \eta_i} = \frac{d}{dt}\frac{\partial (\bar{L}/N)}{\partial (\eta_i)_t},
$$
we can derive the system of evolution equations for parameters 
{$\eta_i = a, b,\zeta, k,\phi$} of the solution (\ref{ansatz}).
Variations on $\zeta$, $k$, $b$ and $a$, respectively, give the following equations
\begin{equation}
k_t =  -\omega^2\zeta - V_0 \frac{\partial G}{\partial\zeta},
\label{var_zeta}
\end{equation}
\begin{equation}
\zeta_t = k, \label{var_k}
\end{equation}
\begin{equation}
a_t = 2ab, \label{var_b}
\end{equation}
\begin{equation}
b_t = \frac{1}{2a^4} - 2b^2 + \frac{qN}{2\sqrt{2\pi}a^3} - \frac{g {N}}{2\sqrt{2\pi}a}\frac{\partial}{\partial a}\left(\frac{1}{a}F\right) -\frac{1}{2}\omega^2 -
\frac{V_0}{a} \frac{\partial G}{\partial a}.\label{var_a}
\end{equation}
{
The equation for the phase $\phi$ is decoupled from the system, and we did not write it here.}
Finally, from (\ref{avL}) we obtain  the system of equations for the soliton width and the center of mass
\begin{equation}
a_{tt} = -\frac{\partial U_a}{\partial a}, \quad  \zeta_{tt}= -\frac{\partial U_{\zeta}}{\partial\zeta},
\end{equation}
where
\begin{eqnarray}
U_a &=&\frac{1}{2a^2} + \frac{N}{\sqrt{2\pi}a}\left(q+gF\right) + \frac{1}{2}\omega^2 a^2 + 2V_0 G,\label{Ua}\\
U_{\zeta} &=& \frac{1}{2}\omega^2 \zeta^2  + V_0 G.
\label{Uz}
\end{eqnarray}
Thus the evolution is described by the dynamics of two coupled  nonlinear oscillators.
From the first equation we can find the fixed point for the soliton width $a_{VA}$ and from the second we can calculate the effective potential $U_{\zeta}(a_{VA})$ for the center of mass of the soliton. 
{
The description of the dipolar soliton dynamics by the system of two coupled nonlinear oscillators was applied successively, for example, in the work \cite{Nath} where soliton-soliton scattering in the dipolar BEC placed in unconnected layers was considered.This method was also used to observe oscillations of the solitons profile in the quasi-1D dipolar BEC in Ref.\cite{Young}.
We also want to note that the Gaussian trial function gives good results for describing the stationary states and dynamics in quasi-1D and -2D geometries (see for example \cite{Nath,Young}). In some cases, for 2D geometries, when the purely dipolar BEC profile (for parameters unstable to collapse)  has blood-cell forms,  this ansatz failed, and it is necessary to choose the trial function as the sum of Gaussian profiles \cite{Ronen,Rau}. In our case the failure of the ansatz can occur when the pulse is splitted into two or more parts by the defect.}


\section{Numerical results}

\subsection{Stationary modes analysis}

We start with the case of competing nonlinearities ($q\cdot g<0$), namely we consider an attractive local nonlinearity, $q<1$, and a repulsive DD interaction, $g>0$, or vice versa.
By using ansatz $\psi(x,t)=\psi(x)e^{-i\mu t}$ and considering $g(t)=g$ we get the stationary equation
\begin{eqnarray}
&&\mu\psi +\frac{1}{2}\psi_{xx}-q|\psi|^2\psi- V_{tr}(x)\psi
\nonumber\\
&&-g\psi(x)\int_{-\infty}^{+\infty} R(|x-\xi|)\ |\psi (\xi)|^2 \ d\xi  = 0.
\label{gpe_st}
\end{eqnarray}

\begin{figure}[h]
\epsfig{file=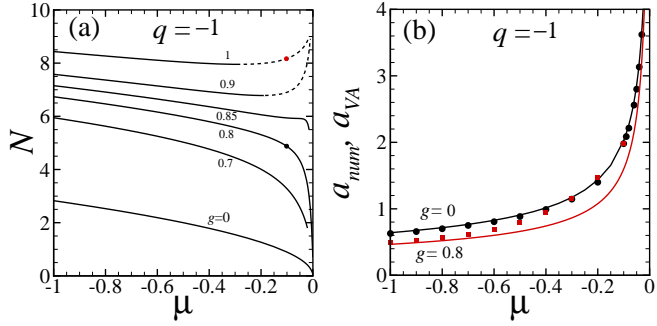,width=9cm}
\caption{In (a) existence curves $N(\mu)$ for attractive local $q=-1$ and for repulsive nonlocal  $0\le g< 1$ interactions. Solid and dashed lines correspond to the stable and unstable regions. In (b) the width calculated from the VA, $a_{VA}$ (solid lines), and from numerical solution of stationary problem, $a_{num}$ (points), are compared.
}
\label{q_1_g}
\end{figure}

Let us first consider stationary solitons without external potential ($\omega=0$) and without defect ($V_0=0$).
In Fig.\ref{q_1_g} and \ref{g_1_q} by solving numerically Eq.(\ref{gpe_st}) (by Newton iteration algorithm) we present the existence curves for families of the solutions with different combination of local and nonlocal terms.
In Fig.\ref{q_1_g}(a) we start with the simple local case ($q=-1$ and $g=0$) for which the norm of the solution  $N$ goes to 0 as the chemical potential $\mu$ goes to 0. By increasing gradually the coefficient $g$ in the nonlocal term we pass through the critical value of $g=g_{cr}\approx 0.87$ for which, at some $\mu=\mu_{cr}$, the existence curve starts to have a local minimum becoming {\it bounded} by critical value $N_{cr}=N(\mu_{cr})$ which means that below $N_{cr}$ the solutions do not exist.
For two branches in Fig.\ref{q_1_g}(a) corresponding to the purely local case ($q=-1$, $g=0$), and with competing local and nonlocal terms, ($q=-1$, $g=0.8$)  we calculated numerically the width of the stationary solutions
\begin{equation}
a_{num}^2=N^{-1}\int_{-\infty}^{\infty} x^2|\psi|^2dx, 
\label{width}
\end{equation}
and compared it with the results of the VA taken from the condition $dU_a/da=0$ (see Eq.(\ref{Ua})).
The results are shown in Fig.\ref{q_1_g}(b). For the purely local case ($g=0$) one observes good agreement between the width calculated numerically and from the VA.
{\it However by increasing the repulsive nonlocal term the discrepancy between $a_{num}$ and $a_{VA}$ starts to grow and for  $g\geq 0.8$ results from the VA for small $|\mu|$ does not agree with the  width of the solution calculated numerically.}

\begin{figure}[h]
\epsfig{file=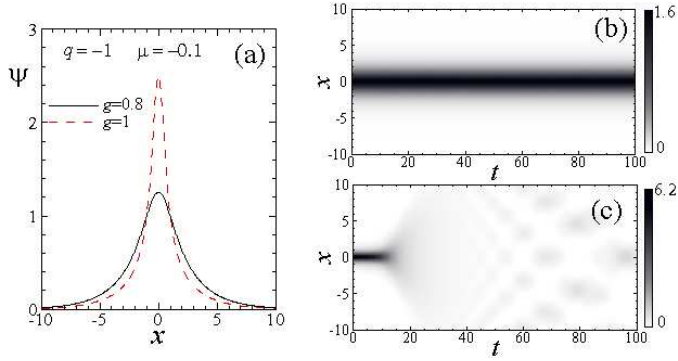,width=9cm}
\caption{ In (a) the profiles of the solution for $q=-1$, $\mu=-0.1$ and different values of coefficient of nonlocal term $g=0.8$ (stable) and $g=1$ (unstable) corresponding to the points in Fig.\ref{q_1_g} (a). In (b) and (c) the density plots of evolution of stable and unstable solutions, checked through direct dynamical simulation of Eq.(\ref{gpe}).
}
\label{dyn}
\end{figure}

In Fig.\ref{dyn} we have checked dynamically the stability of the solutions at the black and red points indicated in Fig.\ref{q_1_g}(a) where the derivative $dN/d\mu$ has different signs.
According to Vakhitov-Kolokolov  stability criterion these two solutions should have different stability.
In Fig.\ref{dyn}(a) the profiles of these two solutions are shown and in the panels (b), (c) their evolutions are presented.
As it is expected the solution with $dN/d\mu<0$ is stable while the solution with $dN/d\mu>0$ is unstable, which is confirmed by direct numerical simulation in Fig.\ref{dyn}(b), (c).

Similar analysis for the existence of the solutions with attractive nonlocal interaction ($g<0$) and with the presence of the local repulsive interaction ($q\geq 0$) is presented in Fig.\ref{g_1_q}.
Now, contrary to the previous case, the norm of the solutions is always unbounded which means that $N\to 0$ as $\mu\to 0$.
Also in this case the width of the solutions calculated numerically from Eq.(\ref{width}) and analytically from the VA are in very good agreement (see Fig.\ref{g_1_q}(b)).

\begin{figure}[h]
\epsfig{file=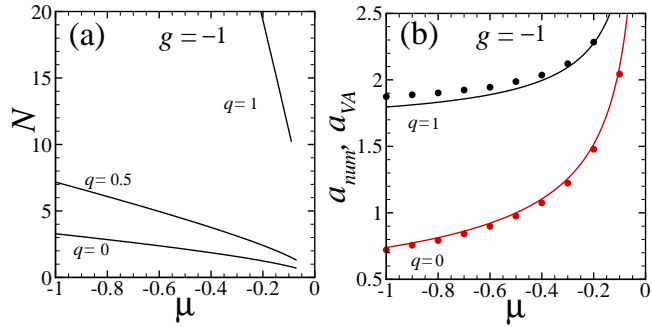,width=9cm}
\caption{The same as in Fig.\ref{q_1_g} for attractive nonlocal interaction, $g=-1$, and for repulsive local interaction $0\le q< 1$.
}
\label{g_1_q}
\end{figure}


\section{Soliton dynamics in the trap and interaction with localized defect}

Now we consider the dynamics of the soliton with local and nonlocal interactions in the presence of the parabolic potential $V_{\omega}$ and the localized defect $V_d$.
Since we are interested in competing local and nonlocal nonlinearities (condition $q\cdot g<0$) in the following we will consider two possible combinations  separately: i) $q<0$, $g\geq 0$; and ii)  $q\geq 0$, $g< 0$.

\subsection{The case $q<0$, $g\geq 0$}

\begin{figure}[h]
\epsfig{file=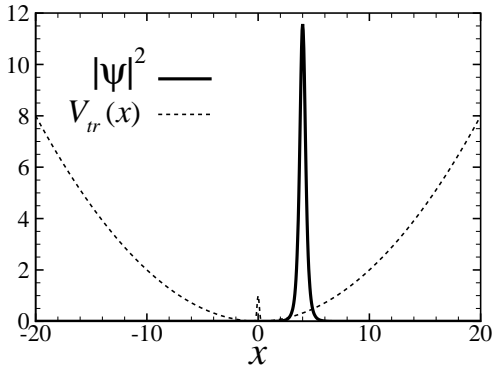,width=8cm}
\caption{Initial profile of the soliton $|\psi(x_0=4, t=0)|$ (solid line) in the parabolic trap $V_{\omega}$ with the defect $V_{d}$ (dashed line). The parameters are: $\mu=-1$, $q=-1$, $g=1$, $V_0=1$, $\ell =0.1$, $\omega=0.2$.
}
\label{fig1}
\end{figure}

Let us first concentrate on the case	of attractive local ($q=-1$) and repulsive nonlocal ($g=1$) interactions when the existence curve is bounded by the critical norm $N_{cr}$, which occurs above $g\approx 0.87$.
In Fig.\ref{fig1} the initial profile of the unperturbed stable soliton taken from the existence curve in Fig.\ref{q_1_g} at $\mu=-1$ and shifted to the position $x_0=4$, as well as the corresponding external parabolic potential with the defect are shown.
The shift from the center is used to get oscillations of the soliton in the defect-free parabolic potential.

\begin{figure}[h]
\epsfig{file=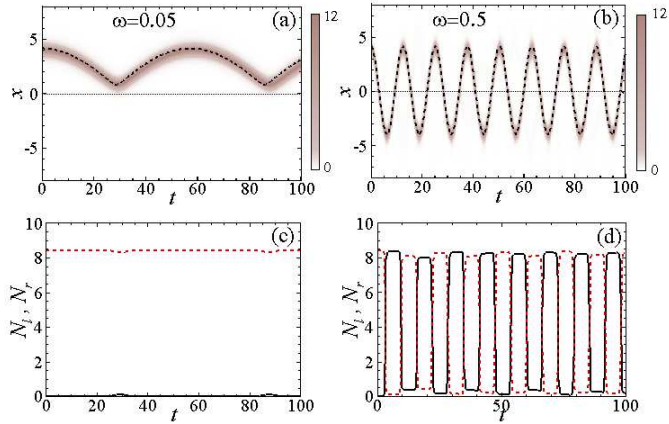,width=9cm}
\caption{In (a), (b) the density plot of the oscillations of the soliton in parabolic potential in the presence of the defect. The strength of the parabolic potential in (a) $\omega=0.05$ and in (b) $\omega=0.5$. Initial profile of the soliton is shifted to $x_0=4$. The dashed black lines correspond to the trajectories of the soliton calculated from the VA (\ref{Ua}), (\ref{Uz}).
Other parameters are: $\mu=-1$, $q=-1$, $g=1$, $V_0=1$, $\ell =0.1$. The dotted line in (a), (b) shows the position of the defect.
In (c) and (d) the evolutions of the norms to the right, $N_r$ (solid black),  and to the left, $N_l$ (dashed red), from the defect are shown corresponding to the cases (a) and (b), respectively.
}
\label{fig2}
\end{figure}

As soon as the soliton is shifted from the minimum of the parabolic potential it starts to accelerate to the center and the soliton velocity at the center, $v_c$, will depend on the strength of the parabolic potential $\omega$ and magnitude of the shift $x_0$ according to $v_c=\omega x_0$.
In this way by changing the strength of the trap with fixed $x_0$ we can control the soliton velocity in the process of the interaction of the soliton with the defect.
In the defect-free parabolic trap the frequency of oscillations of the soliton coincides with the trap frequency ($\zeta(t)=x_0\cos(\omega t)$).
By fixing the amplitude $V_0$  and the width $\ell$ of the defect and letting the soliton collide with the defect we found three characteristic regimes of the soliton-defect interaction depending on the incoming velocity. In Fig.\ref{fig2} we present two regimes. The first one corresponds to a relatively small soliton velocity (weak trap) when the soliton is reflected from the defect and becomes "closed" in the semi-space ($x>0$) of the parabolic potential (see Fig.\ref{fig2}(a)). By increasing the soliton velocity by changing the strength of the parabolic potential we found another limiting case when the soliton has enough kinetic energy to pass through the defect (see Fig.\ref{fig2}(b)). To visualize the corresponding evolutions we calculated the norms in the right part and in the left part from the defect as $N_r=\int_0^\infty |\psi|^2 dx$ and $N_l=N-N_r$. The results are shown in Fig.\ref{fig2}(c),(d) which confirm the above mentioned soliton behavior. 

We also compared direct numerical calculation of the soliton dynamics with the results obtained from the VA. In Fig.\ref{fig2}(a),(b) by dashes lines we present the trajectories of the oscillations of the center of mass of the soliton calculated from Eqs.(\ref{Ua}), (\ref{Uz}).    

\begin{figure}[h]
\epsfig{file=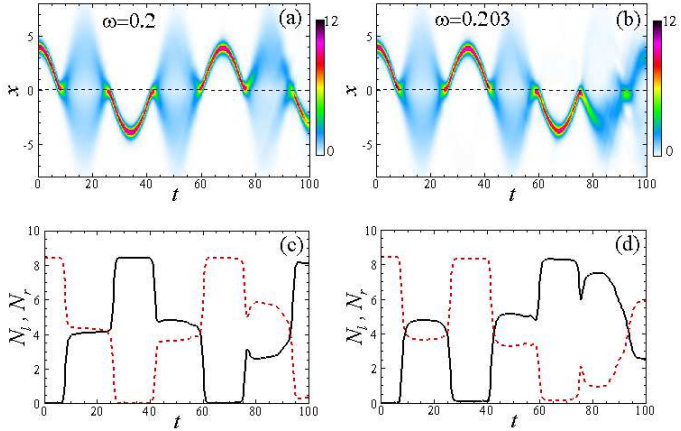,width=9cm}
\caption{The same as in Fig.\ref{fig2} with $\omega=0.2$ in (a) and $\omega=0.203$ in (b).
}
\label{fig3}
\end{figure}

\begin{figure}[ht]
\vspace{1cm}
\epsfig{file=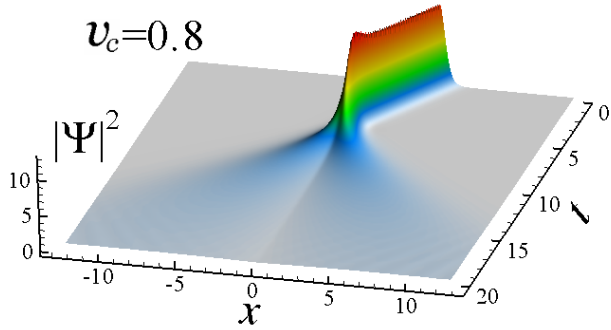,width=8cm}
\caption{Scattering of the soliton with initial velocity $v_c=0.8$ placed at $x_0=4$ in the absence of the parabolic trap ($\omega=0$). Other parameters are the same as in Fig.\ref{fig2}.
}
\label{fig3_v}
\end{figure}

A more interesting effect of the coherent splitting of the soliton by the defect is observed for intermediate values of $\omega$. This is the case when the VA fails and only numerical calculation will be presented.
We found (see Fig.\ref{fig3}) that taking the strength of the parabolic potential around $\omega\approx 0.2$ one can observe splitting of the soliton into two quasi-symmetrical parts. 
This can be seen from the evolution of the norms $N_r$ and $N_l$ in Fig.\ref{fig3}(b), (c).
It should be stressed here that after splitting the soliton lost its solitonic identity as soon as in this case the existence curve has bounded norm $N_{cr}$ and the norm of the each parts is below this critical value ($N_l<N_{cr}$ and $N_r<N_{cr}$).
To confirm this behaviour we switch off the parabolic trap and take a soliton with initial velocity $v_c \approx 0.8$ far from the defect. As it is shown in Fig.\ref{fig3_v}, after interaction with the defect the soliton splits into two packets which transform continuously into linear waves.
In the presence of the trap these linear packets do not escape but they are reflected by the parabolic potential  and return to the center where again produce the initial soliton ("fusion") in the left or right part of the defect and continue this process through several periods.
In Fig.\ref{fig3}(a) one observes that after splitting and returning to the center the soliton continues to move towards the negative $x$ passing completely through the barrier  while in \ref{fig3}(b) soliton after fusion is reflected from barrier.
It should be stressed that very tiny changes in the initial conditions could affect  the condition for the soliton splitting and the dynamics of the soliton after fusion (transmission or reflection scenarios).

\begin{figure}[ht]
\epsfig{file=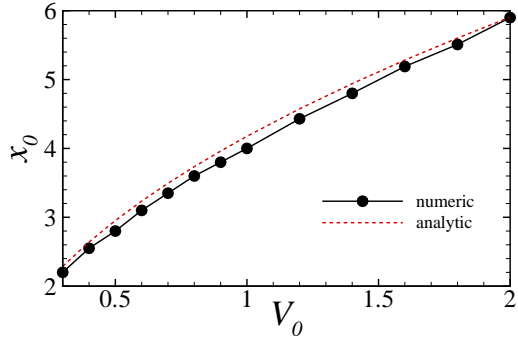,width=8cm}
\caption{Comparison of numerical (line with points) and analytical (dashed line) results using Eq.(\ref{x0}) for the critical coordinate $x_0$ at which one can observe splitting effect ($N_l\approx N_r$). Parameters are: $\omega=0.2$, $\mu=-1$, $q=-1$, $g=1$, $\ell=0.1$. The width of the soliton calculated from the VA is $a_{VA}\approx 0.38$.
}
\label{x0_V0}
\end{figure}

{
In this case the VA can be used to find the condition for soliton splitting. As follows from Eqs.(\ref{var_a}), (\ref{Ua}), when the kinetic energy of the effective particle
$E_{kin}=x_0^2\omega^2/2$ is less then the barrier effective potential height $E_{bar}=V_0G(a_{VA},\ell,\zeta=0)$, a full reflection of the soliton occurs. In the opposite case we have full transmission (compare Figs.\ref{fig2}). Therefore we can assume that when $E_{kin}= E_{bar}$, the partial reflection/transmission should take place. Considering this condition as the soliton splitting condition, we obtain: }
\begin{equation}\label{x0}
x_0 = \frac{1}{\omega}\sqrt{2V_0 G(a_{VA},\ell,\zeta=0)}.
\end{equation}
By comparing the numerical results with the analytical ones taken from (\ref{x0}) one observes excellent agreement (see Fig.\ref{x0_V0}).

\begin{figure}[h]
\epsfig{file=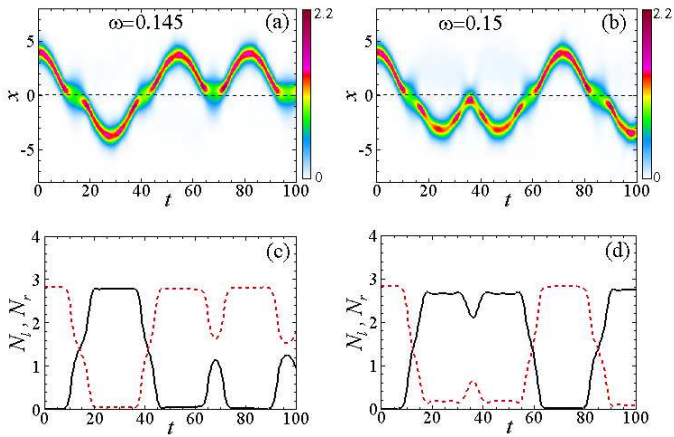,width=9cm}
\caption{Dynamics of the soliton with $q=-1$ and $g=0$ in the parabolic trap with the strengths $\omega=0.145$ (a),(c) and $\omega=0.15$ (b),(d).
}
\label{fig3_g0}
\end{figure}

For smaller strength of the repulsive nonlocal nonlinearity, when the norm of the solution is unbounded,  one observes a rather different picture of splitting of the soliton at the defect (see Figs.\ref{fig3_g0})


Comparing Fig.\ref{fig3} and Fig.\ref{fig3_g0} one can conclude that in the case $g=1$ the soliton transforms into two quasi-identical wave packets in Fig.\ref{fig3}(a),(b) (one also observes this effect looking at $N_{l,r}$ when along a half period $N_l\approx N_r$) while in the case $g=0$ there is no splitting of the soliton. Instead of that it transforms into the defect mode  localized at the defect.

\begin{figure}[h]
\epsfig{file=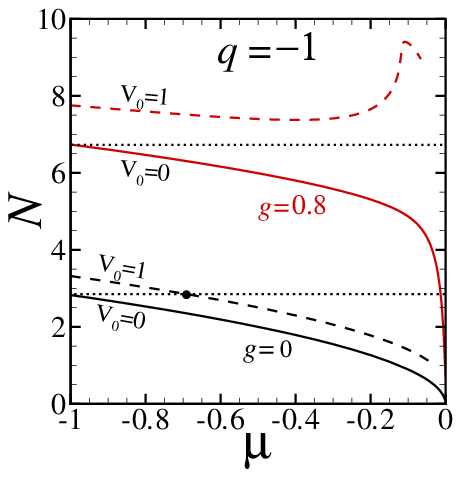,width=7cm}\epsfig{file=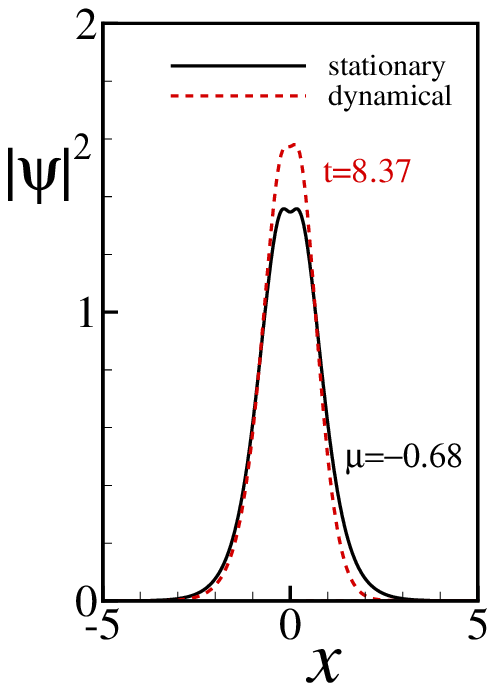,width=4.5cm}
\caption{Left panel: Existence curves without defect $V_0=0$ (solid lines) and with defect $V_0=1$ (dashed lines) for the cases $g=0$ and $g=0.8$.  Right panel: The profiles of the defect modes with the same number of particles calculated form stationary problem (black solid) at the black point in the left panel  with $\mu=-0.68$ and from dynamical equation (dashed red) where initial solution was taken at $\mu=-1$ for $V_0=0$.
}
\label{q_1_g_v}
\end{figure}

This can be verified in Fig.\ref{q_1_g_v} by comparing the existence curves for the case without defect   with curves calculated in the presence of the defect.
As one can see in Fig.\ref{q_1_g_v} by switching the defect  on the existence curves go upper and what is essential is that the critical value of $g$ at which the existence curves become bounded also decreases with presence of the defect. As an example the existence curve for $g=0.8$ in the defect-free case, $V_0=0$,  is unbounded while in the presence of the defect with $V_0=1$, the existence curve becomes bounded.
Considering the case $g=0$ the soliton at $\mu=-1$ and $V_0=0$  has the same number of particles  as the defect mode at $\mu=-0.68$  and $V_0=1$ (see the lower horizontal dotted line in the left panel of Fig.\ref{q_1_g_v}).
To check this we calculated the defect mode for $\mu=-0.68$ in the presence of the defect and compared it with the defect mode obtained by direct dynamical calculation where the initial soliton for $\mu=-1$ and $V_0=0$ has the same number of particles as the defect mode.
In the right  panel of Fig.\ref{q_1_g_v} we compare these two defect modes  (time $t=8.37$ corresponds to the instant when $N_r\approx N_l$ during interaction of the soliton with the defect).

In the case $g=0.8$ the situation is different. The number of particles in the soliton at $\mu=-1$ and $V_0=0$ is below the critical number of particles, $N_{cr}$, needed to generate the defect mode in the presence of the defect (there is no intersection of the horizontal dotted with the upper dashed line).  In this case one can observe splitting of the soliton into two linear wave packets after interaction with the defect.

\paragraph{Localization-delocalization transition governed by the nonlocal term.}

\begin{figure}[h]
\epsfig{file=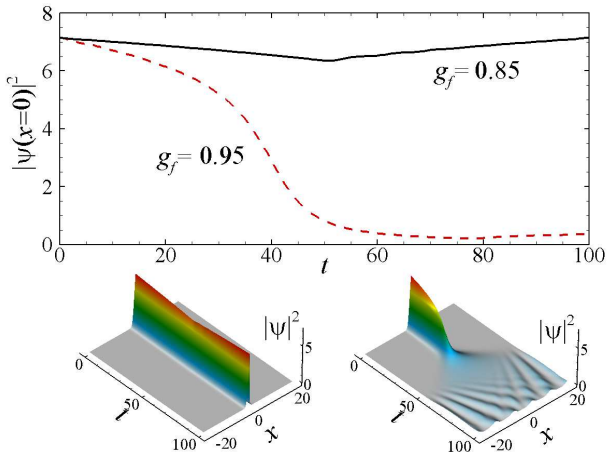,width=9cm}
\caption{Localization-delocalization transition managed by nonlocality. In (a) the amplitude of the density in the center $|\psi(x=0)|^2$ for two different values of  $g_f$: black solid line $g_f=0.85$ and red dashed line $g_f=0.95$. In (b) and (d) the density plot of the corresponding dynamics. Other parameters: $\mu =-1$, $q=-1$, $g_i=0.8$, $\omega=0$, $V_0=0$.
}
\label{deloc}
\end{figure}

Let us consider a linear variation in time  of the strength of the nonlinear nonlocal coefficient of the form
\begin{equation}
g(t)=g_f+(g_i-g_f)|1-2t/t_f|
\label{gt}
\end{equation}
where $g_i= g(t = 0)$ and $g_f = g(t = t_f /2)$ correspond
to the initial strength of the nonlocal coefficient and its value at the turning point,
$t = t_f /2$, with the return to the initial value at $t = t_f$.
Tuning the dipolar interactions in quantum gases can be achieved using for example time dependent control of the anisotropy of dipolar interactions  suggested in \cite{Giovanazzi}.
{As it was shown in Section 4.1, in the presence of the attractive local and the repulsive nonlocal interaction (see Fig.\ref{q_1_g}) one can have a transition between unbounded and bounded cases of the existence curves. Using a time dependent nonlocal coefficient $g(t)$}
in Fig. \ref{deloc} we present evolutions of the profile of the soliton density with $|g_f|<|g_{cr}|$ and $|g_f|>|g_{cr}|$ (here $g_{cr}$ corresponds to the case when the norm of the initial soliton coincides with the norm at the global minimum of the existence curve $N_{cr}$). As one can see in the former case there is no delocalization transition and the solution remains localized, while in the latter case at some instant the norm of the solution becomes smaller then the critical norm $N_{cr}$ and the norm of the solution any more pretends to the existence curve and therefore the solution decays into linear waves.
Returning to the initial value $g=g_i$ in the first case the solution reconstructs its initial form while in the second case it remains delocalized.

\subsection{The case $g<0$, $q\geq 0$}

Now let us consider the opposite situation when the solitonic structure is supported by an attractive nonlocal interaction $g<0$ and the local interaction is repulsive, $q>0$.

\begin{figure}[h]
\epsfig{file=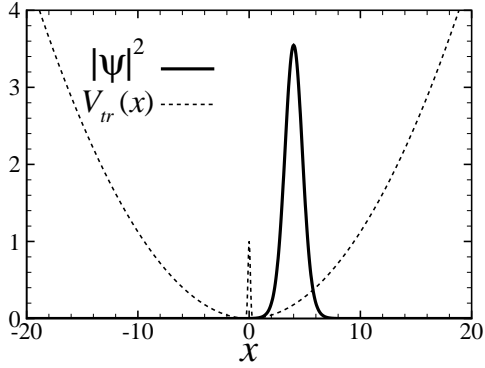,width=8cm}
\caption{Initial profile of the soliton $|\psi(x_0=4, t=0)|$ in the parabolic trap $V_{\omega}$ with the defect $V_{d}$. The parameters are: $\mu=-1$, $q=1$, $g=-1$, $V_0=1$, $\ell =0.1$, $\omega=0.15$.
}
\label{fig11}
\end{figure}

In Fig.\ref{fig11} the corresponding initial profile of the nonlocal soliton in the parabolic trap with defect is shown.
As in the previous case the shift from the center of the trap is needed to observe oscillations of the solitons and eventual interaction with defect placed in the center of the trap.

\begin{figure}[ht]
\epsfig{file=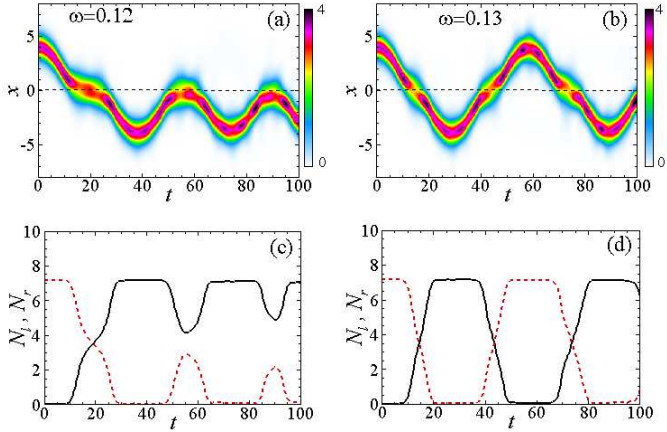,width=9cm}
\caption{Dynamics of the dipolar soliton with $q=0.5$ and $g=-1$ in the parabolic trap with presence of the defect $V_0=1$, $\ell=0.1$.In (a), (c) the evolution of the density plots are shown for $\omega=0.14$ in (a) and $\omega=15$ in (b) with corresponding dynamics of the norms $N_r$ and $N_l$.
}
\label{q-05}
\end{figure}

Two simple scenarios of the interaction of the soliton with the defect in the case of a strong parabolic trap are shown in Fig.\ref{q-05}(a), (c). In the first case the incoming soliton has enough kinetic energy to go through the defect. In this interaction the soliton lost part of the energy and therefore became locked by the defect in the semi-space $x<0$. For a stronger trap the soliton has enough kinetic energy to go through the barrier several times (see Fig.\ref{q-05}(b), (d)).

\begin{figure}[ht]
\epsfig{file=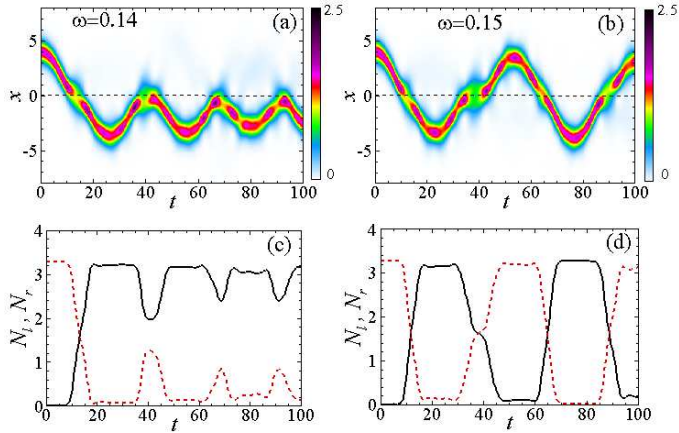,width=9cm}
\caption{ The same as in Fig.\ref{q-05} for the case of the pure dipolar soliton with $q=0$ and $g=-1$. The strength of the parabolic trap in (a), (c) is $\omega=0.14$ and in (b),(d) is $\omega=0.15$.
}
\label{q0}
\end{figure}

A similar picture of the scattering of the pure  dipolar soliton can be observed by decreasing the strength of the local interaction $q$ {to zero} (see Fig.\ref{q0}). As one can see the pure dipolar soliton does not present very robust dynamics, what can be explained by the weak force of the nonlocal interaction.

\section{Conclusion}
In conclusion, we have investigated the dynamics of bright solitons in a dipolar condensate loaded into a parabolic trap with a barrier potential at the center. Using  a variational approach and  scattering theory, we have studied the reflection, transmission and splitting of bright solitons on the defect in the presence of the trap potential. We explore the different sets of parameters changing the relative strength of the local and nonlocal (dipolar) interactions. The case when the local nonlinearity dominates is close to the settings investigated early \cite{Holmes,Holmer,Cao} and well understood now. The intermediate cases when $q \sim g$ and the pure dipolar nonlinearity $q=0$ are considered here.

We show that coherent splitting of the dipolar bright soliton exists. The splitting parts are reflected by  the trap and recombine into a single soliton, performing a few oscillations under the trap potential. The condition for soliton splitting is derived, which is in excellent agreement with numerical simulations of the full nonlocal Gross-Pitaevskii equation.

Localization-delocalization phenomena governed by the variation of dipolar interactions in time has been studied for the cases $|g_f| < |g_{cr}|$ and $|g_f| > |g_{cr}|$.
The reconstruction of the soliton to its initial form is observed for the former case, while in the latter case the soliton transforms into the linear waves.  

The obtained results can be interesting for the study of soliton dynamics in the case of very small atomic scattering length when the dipolar interactions effects start to play a dominant role. Several experiments were performed recently in $^7$Li with a very small atomic scattering length
tuned via the broad $|1,1>$ Feshbach resonance with the number of atoms in the soliton of the order of $N \sim 2\cdot 10^5$ \cite{Dyke,Pollack}. The barrier potential was generated by a near-resonant cylindrically focused laser beam.   It can be useful for the design of matter--wave beamsplitters and matter--wave interferometers using the fusion of the solitons.

The related problem of quantum scattering of solitons by the potential {\it well} for BEC with local interactions has been studied in \cite{Brand,AGT,Baizakov}. For the case of the dipolar solitons this problem will be considered separately.

\section*{Acknowledgments}
{FKA was  supported by  the 7th European Community Framework Programme under the
grant PIIF-GA-2009-236099 (NOMATOS).}

\end{document}